\documentclass[aps,tightenlines,showpacs]{revtex4}
\usepackage{graphicx}
\usepackage{float}
\usepackage[T1]{fontenc}
\usepackage[latin1]{inputenc}
\usepackage{amssymb}
\include{epsf}
\epsfverbosetrue

\newcommand{\bea}{\begin{eqnarray}}
\newcommand{\eea}{\end{eqnarray}}
\newcommand{\Qsl}{Q\hspace{-.22cm}/~}

\newcommand{\rhatsl}{\hat r \hspace*{-.17cm}/}

\makeatother
\begin{document}

\title{The fermion mass at next-to-leading order in the HTL effective theory}

\author{M.E. Carrington, A. Gynther}

\affiliation{Department of Physics, Brandon University,
Brandon, Manitoba, R7A 6A9 Canada\\
and Winnipeg Institute for Theoretical Physics,
Winnipeg, Manitoba, Canada}

\author{D. Pickering}

\affiliation{Department of Mathematics, Brandon University,
Brandon, Manitoba, R7A 6A9 Canada}

\begin{abstract}
The calculation of the real part of a quasi-particle dispersion relation at next-to-leading order in the hard thermal loop effective theory is a very difficult problem. Even though the hard thermal loop effective theory is almost 20 years old, there is only one next-to-leading order calculation of the real part of a quasi-particle dispersion relation in the literature \cite{SCH}. In this paper, we calculate the fermion mass in QED and QCD at next-to-leading order. 
For QED the result is $M=e T/\sqrt{8}\,(1-(1.427\pm 0.02)e/4\pi)$ and for QCD with $N_f = 2$ and $N_c = 3$ we obtain $M=gT/\sqrt{6}\,(1+(1.867\pm 0.02)g/4\pi)$.
\end{abstract}


\maketitle

\section{Introduction}

It is well known that the behaviour of an elementary particle becomes modified when the particle propagates in a medium. The particles become ``dressed'' by their interaction with the medium, and one speaks of collective modes, or quasi-particles. 
One studies these collective modes by looking at the corresponding thermal propagators. The behaviour of the quasi-particles is deduced from the analytic structure of the propagator. 
In \cite{KKR} it was shown from general principles that the singularity structure of certain components of gauge and matter propagators are gauge-independent, when all contributions of a given order are systematically taken into account. 

The calculation of dispersion relations for soft quantities at next-to-leading order (NLO) in the hard thermal loop (HTL) effective theory is notoriously difficult. There are several calculations of damping rates at NLO. The soft static gluon damping rate was calculated in Ref. \cite{BPgluondamp}. The damping rate of a soft static quark was calculated in Ref. \cite{KKM,BPquarkdamp,MC}. Calculations of masses and oscillation frequencies require the real part of the dispersion relation, which is considerably more difficult to obtain. 
There is only one complete calculation in the literature of the real part of a quasi-particle dispersion relation at NLO: the pure glue plasma frequency in the long wavelength limit was calculated by Schulz \cite{SCH}. In this paper we calculate the fermion mass, in QED and QCD, at NLO. 
It is straightforward to obtain the result for QCD from the corresponding result for QED by adjusting the HTL masses and including an overall factor $C_F$ in the quark self energy. 

For soft static electrons with momenta $Q_\mu=(q_0\sim e T,\vec q=0)$, the mass and damping rate of the quasi-particle are obtained from the solution of the equation
\bea
\label{dispX}
{\rm det}\,(\Qsl-{\bf \Sigma}_{ret}(Q))\Big|_{q_0=M-i\gamma} & = & 0.
\eea
The fermion self-energy can be decomposed in the usual way:
\bea
{\bf\Sigma}_{ret} & = & \gamma^0\Sigma^{(0)}_{ret} + \vec{\gamma}\cdot\hat{q}\Sigma^{(i)}_{ret}.
\eea
Since we have taken $q=0,$ the only non-zero component is 
$\Sigma_{ret}^{(0)}(q_0) = {\rm Tr}\,(\gamma^0 \Sigma_{ret}(Q))/4$. 
From now on, we suppress the superscript `$(0)$' to simplify the notation. In addition, we suppress throughout the subscript `$ret$' indicating the retarded component of the self energy. Using this notation we write the dispersion relation as
\bea
\label{disp}
q_0-\Sigma(q_0)\Big|_{q_0=M-i\gamma} & = & 0.
\eea

At leading order (LO), the self energy is given by the familiar HTL result,
\bea
\label{HTL}
{\rm Re}\,\Sigma_{\rm HTL} & = & \frac{e^2T^2}{8q_0},\,{\rm Im}\,\Sigma_{\rm HTL} = 0.
\eea
Substituting (\ref{HTL}) into  (\ref{disp}) we obtain the leading order results for the mass and damping rate:
\bea
\label{lo}
m_f:=M^{(0)} & = & \frac{e T}{\sqrt{8}},\,\gamma^{(0)} = 0.
\eea

We are interested in obtaining NLO corrections to these results.
% which we define as corrections of order $e(e T)$. 
To obtain these NLO corrections, we expand the dispersion relation around the lowest order (LO) solution in Eqn. (\ref{lo}), keeping contributions to linear order in NLO quantities. The resulting equations are particularly simple because of the fact that the imaginary part of the LO HTL self energy is zero. The real and imaginary parts of the dispersion relation give
\bea
M^{(1)}-M^{(1)}{\rm Re}\,\Sigma_{\rm HTL}^\prime[M^{(0)}]-{\rm Re}\,\Sigma_{NLO}\big[M^{(0)}\big] & = & 0,\\[2mm]
\gamma^{(1)}-\gamma^{(1)}{\rm Re}\,\Sigma_{\rm HTL}^\prime\big[M^{(0)}\big] + {\rm Im}\,\Sigma_{NLO}\big[M^{(0)}\big] & = & 0.\nonumber
\eea
Using (\ref{HTL}) and (\ref{lo}) we obtain
\bea
\label{NLO}
M^{(1)} & = & \frac{1}{2}{\rm Re}\,\Sigma_{NLO}\big[M^{(0)}\big],\,\gamma^{(1)}=-\frac{1}{2}{\rm Im}\,\Sigma_{NLO}\big[M^{(0)}\big].
\eea

To obtain the NLO mass and damping rate from (\ref{NLO}) we must calculate the NLO self energy. 
The original paper by Braaten and Pisarski \cite{BP} identified three potential contributions. They are: (1) corrections to the LO result for the 1-loop diagram obtained by expanding to next  order in the ratio of the soft external momentum to the hard loop momentum; (2) contributions to the 2-loop diagrams from the region of the phase space that corresponds to both loops hard; and (3) contributions to 1-loop diagrams with soft loop momentum, and all propagators and vertices replaced with HTL effective ones. The power counting arguments of Braaten and Pisarski refer to the maximum possible contribution from each type of term. The actual contribution may be lower order for kinematical reasons, or because of some cancellation between different integrals. 

\begin{figure}[H]
\begin{center}
\includegraphics[width=10cm]{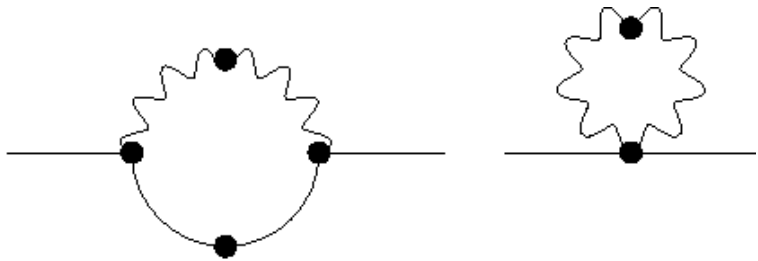}
\end{center}
\caption{The diagrams that contribute to the self-energy up to NLO. Wavy lines indicate HTL photons and solid lines are HTL fermions. The vertices are defined in Eqn. (\ref{vertex}).} 
\label{KKMdiag2}
\end{figure}

The full NLO contribution is contained in the dressed 1-loop diagrams shown in Fig. \ref{KKMdiag2}, where the dots on the vertices indicate the sum of the bare vertex and the HTL vertex. These diagrams contain all of the contributions identified by Braaten and Pisarski, if the loop momentum is integrated over the full range from zero to infinity. The integral will also contain subleading contributions that are suppressed by powers of coupling. 

%%%%%%%%%%%%%%%%%%%
\par
%%%%%%%%%%%%%%%%%%%%

It has been demonstrated that the NLO contribution to the dispersion relation from the diagrams in Fig. \ref{KKMdiag2} is gauge invariant. This result is obtained by using the fact that the HTL vertices and propagators satisfy the usual Ward identities. One finds that the gauge dependent contribution to the NLO fermion self-energy is proportional to an integral times the square of the inverse propagator $S^{(-1)}(Q) = \Qsl - \Sigma_{\rm HTL}(Q),$ which vanishes on the mass shell. In \cite{BKS} it was pointed out that a straightforward evaluation of the integral produces mass-shell singularities that cancel the contributions from the two inverse propagators, and give a finite gauge dependent contribution to the damping rate. This problem was resolved by Rebhan \cite{Tonycomment} who showed that the integral must be regulated before the mass shell is approached. Using this procedure one finds that the position of the pole is gauge independent, and the gauge dependence occurs only in the unphysical residue.  

The integral that corresponds to the diagrams in Fig. \ref{KKMdiag2} has the general form
\bea
\label{int1}
\Sigma(q_0) & = & e^2 \int dp_0 \int dp\, {\cal F}(q_0,p_0,p,m_f,n_b(p_0),n_f(p_0))\Big|_{q_0=m_f}.
\eea
The factor of $e^2$ in front of the integral is the explicit factor coming from the two vertices. The integrand is obtained by combining HTL propagators and HTL vertex functions, and thermal distribution functions (defined in (\ref{therm})). The HTL propagators and vertices depend on the 4-momenta and the HTL fermion mass $m_f$. As explained above, we substitute the LO result $q_0=m_f$ in order to extract the NLO contribution. 

We begin by noting that, if the integral in Eqn. (\ref{int1}) is dominated by the part of phase space that corresponds to $p$-soft, we can expand the thermal distribution functions and use: $n_b(p) \to T/p$ and $n_f(p) \to 0$. After expanding the distribution functions, one can extract a factor of the temperature, and scale all remaining variables by the LO mass $m_f$. The result has the form
\bea
\label{int2}
\Sigma_{\rm NLO} = e^2 T ~\cdot~{\cal I}\,,
\eea
where ${\cal I}$ is a dimensionless integral that can be calculated numerically. 

For the imaginary part of the self energy, we have explicitly calculated the 2-loop contributions and checked that the integral in (\ref{int1}) is dominated by the $p$-soft region of the phase space. The numerical calculation of the integral represented in (\ref{int2}) has been done previously \cite{KKM,BPquarkdamp,MC}. The result is \footnote{Note that in \cite{MC} the damping rate was defined with an extra factor of 2.}:
\bea
\gamma_{\rm QED} & = & \frac{e^2 T}{4\pi}\;\cdot \;(1.35), \\
\gamma_{\rm QCD} & = & \frac{g^2 T C_F}{4\pi}\;\cdot \;(1.41) ~~~~{\rm for}~~N_c=3,~N_f=2,~C_F=4/3.\nonumber
\eea

In this paper we calculate the real part of the self energy by evaluating numerically the integrals that correspond to the diagrams in Fig. \ref{KKMdiag2}, without expanding the distribution functions. We extract the numerical coefficients of the NLO terms by extrapolating to small values of the coupling constant. The result of this computation is
\bea\label{result}
M_{\rm QED} & = & \frac{e T}{\sqrt{8}}\left[1 - (1.427\pm 0.02)\frac{e}{4\pi}\right] +\mathcal{O}(e^3T),\\[2mm]
M_{\rm QCD} & = & \frac{g T}{\sqrt{6}}\left[1 + (1.867\pm 0.02)\frac{g}{4\pi}\right]+ \mathcal{O}(g^3T),\hspace{1cm} (N_f = 2,\, N_c = 3). \nonumber
\eea

%{\bf Here some discussion about the relation of the $e^3\ln 1/e$ analytic result to the present numeric result}

\section{Notation}
In this section we define our notation and give the integrals that determine the real part of the fermion self energy at NLO. 
We use
\bea
\{\gamma_\mu,\gamma_\nu\} & = & 2g_{\mu\nu},\,g_{\mu\nu} = {\rm diag}\;(1,-1,-1,-1).\nonumber
\eea
The thermal distribution functions are defined as
\bea
\label{therm}
n_b(p) & = & \frac{1}{e^{\beta p}-1},\,n_f(p) = \frac{1}{e^{\beta p}+1},\,N_B(p) = 1+2n_b(p),\,N_F(p) = 1-2n_f(p).
\eea
In this paper we are only interested in thermal effects and consequently we ignore zero temperature pieces of the self-energy. We use capital letters to denote 4-momenta: $K = (k_0,\vec{k})$. We take the external momentum to be $Q = (q_0,\vec{0})$ and the loop momentum is $P = (p_0,\vec{p})$. We write $R=P+Q$ so that we have $\vec{r}=\vec{p}$. Retarded propagators and self energies are obtained from $p_0 \to p_0+i \epsilon$ and advanced functions from $p_0 \to p_0-i \epsilon$. 
In QED, to leading order the fermion and photon thermal masses are 
\bea
m_f^2 & = & \frac{e^2 T^2}{8},\,m_G^2 = \frac{e^2T^2}{6},
\eea
and in QCD 
\bea
m_f^2 & = & \frac{g^2 T^2}{8}C_F,\,m_G^2 = \frac{g^2T^2}{6}\left(N_c+\frac{1}{2}N_f\right),\,C_F=\frac{N_c^2-1}{2N_c},
\eea
where $N_c$ is the number of colours and $N_f$ is the number of flavours. 

The HTL self energy is
\bea
\Sigma_{\rm HTL}(Q) & = & \frac{m_f^2}{q} {\cal Q}_0(q_0,q),\, {\cal Q}_0(q_0,q) = \frac{1}{2} \ln \left(\frac{q_0+q}{q_0-q}\right),
\eea
and we define
\bea
\Sigma_{\rm HTL}(d)(Q) & = & 2i\,\mathrm{Im}\,\Sigma_{\rm HTL}^{ret}(Q),~~\Sigma_{\rm HTL}(s)(Q) = 2\mathrm{Re}\,\Sigma_{\rm HTL}^{ret}(Q).
\eea
The  HTL fermion propagators are written as
\bea
\label{ferm2}
S(R) & = &\frac{1}{2}(S_+(R)(\gamma_0-\rhatsl)+\frac{1}{2}S_-(R)(\gamma_0+\rhatsl)),  \\
S_+(R) & = & -\frac{2 r^2}{2 r \left(m_f^2+r
   \left(r-r_0\right)\right)+\ln \left(\frac{r_0+r}{r_0-r}\right) \left(r-r_0\right) m_f^2}, \nonumber\\
S_-(R) & = & \frac{2 r^2}{2 r \left(m_f^2+r \left(r+r_0\right)\right)-\ln \left(\frac{r_0+r}{r_0-r}\right) m_f^2
   \left(r+r_0\right)}.\nonumber
\eea
We use the covariant gauge and write the photon propagator in terms of transverse and longtitudinal components (recall that $p=r$),
\bea
\label{photon-prop}
D_{\mu\nu}(P) & = & P_{\mu\nu}^T D_T(P) +P_{\mu\nu}^L \frac{p^2}{P^2}D_L(P),\\[2mm]
D_T(P) & = &\frac{1}{P^2-G(p_0,r)},~~D_L(P)=\frac{P^2}{r^2}\frac{1}{P^2-F(p_0,r)},\nonumber\\[2mm]
G(p_0,r) & = & \frac{1}{r^2}\left(1-\frac{{\cal Q}_0\left(p_0,r\right) p_0}{r}\right) P^2 m_G^2+m_G^2,\nonumber\\[2mm]
F(p_0,r) & = & -\frac{1}{r^2}2 m_G^2 \left(1-\frac{{\cal Q}_0\left(p_0,r\right) p_0}{r}\right) P^2.\nonumber
%&& P_{\mu\nu}^T = g_{\mu  \nu }-\frac{Q^{\mu } Q^{\nu }}{q_0^2}+\frac{1}{r^2}\left(P^{\mu }-\frac{p_0 Q^{\mu }}{q_0}\right) \left(P^{\nu}-\frac{p_0 Q^{\nu }}{q_0}\right)\nonumber\\[2mm]
%&& P_{\mu\nu}^L =-\frac{P^{\mu } P^{\nu }}{P^2}+\frac{Q^{\mu } Q^{\nu }}{q_0^2}-\frac{1}{r^2}\left(P^{\mu }-\frac{p_0 Q^{\mu}}{q_0}\right) \left(P^{\nu }-\frac{p_0 Q^{\nu }}{q_0}\right) \nonumber
\eea
Furthermore, we define the discontinuities and the principle parts as
\bea
d_\pm(R) & = & 2i\mathrm{Im}\, S^\pm_{ret}(R),~~{\cal P}_\pm (R) = \mathrm{Re}\,S^\pm_{ret}(R),\\
d_{T/L}(P) & = & 2i\mathrm{Im}\,D^{T/L}_{ret}(P),~~{\cal P}_{T/L} (P) = \mathrm{Re}\,D^{T/L}_{ret}.\nonumber
\eea   
Here, each discontinuity contains a pole contribution and a cut contribution:
\bea
d_{T/L}(K) & = &-2\pi i \sum_{n=\pm 1} n\, Z_{T/L}(\omega_{T/L}(k),k)\; \delta(k_0-n \,\omega_{T/L}(k))-2\pi i \beta_{T/L}(k_0,k), \\
d_\pm(K) & = & -2\pi i\;Z(k_0,k)\;\big[\delta(k_0-\omega_\pm(k))+\delta(k_0+\omega_\mp(k))\big]-2\pi i \beta_\pm(k_0,k).\nonumber
\eea
Expressions for the functions $\{\beta_T,~\beta_L\,~\beta_+,~\beta_-\}$, and the equations from which $\{\omega_T,~\omega_L\,~\omega_+,~\omega_-\}$, are obtained can be found using Eqns. (\ref{ferm2}) and (\ref{photon-prop}). They are also given in the appendix of \cite{MC}. 

\section{Integrands}

At zero temperature, the integral corresponding to the diagrams in Fig. \ref{KKMdiag2} can be written as
\bea
\Sigma(Q) & = & -i \,e^2\,\int dP\big(\Gamma_\mu(Q,P+Q) S(P+Q) \Gamma_\nu(P+Q,Q)\;D^{\mu\nu}(P)+M_{\mu\nu}(Q,P,-P,Q)\,D_{\mu\nu}(P)\big),
\eea
where $\int dP:=\int dp_0\int d^3p$, and $i{\cal S}$ and $-iD_{\mu\nu}$ correspond to the electron and photon lines respectively. We need to obtain the corresponding integral at finite temperature. We work in the Keldysh representation of the real time formalism. The method we use to sum over Keldysh indices is described in \cite{mcTF}. 
The vertices $\Gamma$ and $M$ are defined in Eqn. (\ref{vertex}) where $P_{\psi {\rm in}}$ indicates the momentum of an incoming fermion, $P_{\psi {\rm out}}$ is the momentum of an outgoing fermion, and $P^\mu_{\gamma{\rm in}}$ is the momentum of an incoming photon,
\bea
\label{vertex}
\Gamma_\mu(P_{\psi {\rm in}},P_{\psi {\rm out}}) & = & \gamma_\mu+\Gamma^{\rm HTL}_\mu(P_{\psi {\rm in}},P_{\psi {\rm out}}),\\[2mm]
M_{\mu\nu}(P_{\psi {\rm in}},P^\mu_{\gamma{\rm in}},P^\nu_{\gamma{\rm in}},P_{\psi {\rm out}}) & = & M^{\rm HTL}_{\mu\nu}(P_{\psi {\rm in}},P^\mu_{\gamma{\rm in}},P^\nu_{\gamma{\rm in}},P_{\psi {\rm out}}).\nonumber
\eea

A complete expression for the integrand is derived in Ref. \cite{MC}.  A reasonably simple form is obtained by rewriting the HTL vertices in terms of the self energies, and rearranging the result. There are several tricks that must be used to remove the dependence on the HTL vertices. First, one uses the Kubo-Martin-Schwinger (KMS) conditions for 3- and 4-point functions to obtain an expression that depends only on retarded vertex functions.  A complete list of the KMS conditions for 3- and 4-point functions is found in \cite{mcTF}. For many terms, the HTL Ward identities can be used to replace contractions of HTL vertices with the connecting photon momentum by the appropriate difference of HTL self energies. There are some terms for which one must use explicit results for the HTL 3-point vertex functions. Fortunately, these expressions have a particularly simple form when one of the fermions is not moving.
Using these techniques, all components of the vertices can be written as simple functions of the HTL self-energy. 
These self-energies also appear in the denominators of the HTL fermion propagators. The general strategy is to rearrange terms in the numerators to cancel as many terms as possible with the corresponding factors in the denominators. Significant simplifications occur after combining terms and using the mass shell condition $q_0^2=m_f^2$. 
The imaginary part of the resulting expression, which determines the damping rate at NLO, has been evaluated numerically in Refs. \cite{KKM,BPquarkdamp,MC}. 

The calculation of the real part of the self energy,  which determines the NLO contribution to the mass, is more complicated for several reasons. One problem is that second diagram in Fig. \ref{KKMdiag2} produces pure real tadpole type contributions that can be dropped in the calculation of the imaginary part. There are additional numerical complications that will be discussed in more detail in Sec.~\ref{sec:num}. 
We give below the integrals that need to be calculated to obtain the NLO contribution to the real part of the self energy \cite{MC}. We separate terms that contain different combinations of delta functions, principle parts, and thermal factors. In addition, we define the operator
\bea
\label{Nhat}
\hat N & := & -\frac{i e^2}{32 \pi ^3} \int dp_0\int dr,
\eea
which will be factored out of all expressions. The integrals are
\bea
\label{real-part}
\text{Re}\Sigma(d_L,N_B) & = & \frac{12}{q_0} \;\hat N \;r^2 d_L(p) N_B\left(p_0\right),\\
\text{Re}\Sigma(d_L,{\cal P}_+,N_B) & = & -\frac{2}{q_0^2} \;\hat N \;r^2 \left(-r+q_0+r_0\right)^2 d_L(P) N_B\left(p_0\right)
   {\cal P}_{+}(R),\nonumber\\
\text{Re}\Sigma(d_L,{\cal P}_-,N_B) & = & -\frac{2}{q_0^2} \;\hat N \;r^2 \left(r+q_0+r_0\right)^2 d_L(P) N_B\left(p_0\right)
   {\cal P}_{-}(R), \nonumber\\
\text{Re}\Sigma(d_+,{\cal P}_L,N_F) & = & -\frac{2}{q_0^2}\;\hat N \; r^2 \left(-r+q_0+r_0\right)^2 d_{+}(R)
   N_F\left(r_0\right) {\cal P}_L(P), \nonumber\\
\text{Re}\Sigma(d_-,{\cal P}_L,N_F) & = &-\frac{2}{q_0^2} \;\hat N \;r^2 \left(r+q_0+r_0\right)^2 d_{-}(R)
   N_F\left(r_0\right) {\cal P}_L(P), \nonumber\\
\text{Re}\Sigma(d_T,N_B) & = & -\frac{1}{q_0^2}\;\hat N \; d_T(p) N_B\left(p_0\right) \left(2 \left(-3 r_0 R^2 +4 m_f^2 r_0+8 r^2 q_0+6 P^2 q_0\right)-R^2 \Sigma_{\rm HTL}(s)(R)\right),\nonumber\\
\text{Re}\Sigma({\cal P}_T,N_F) & = & \frac{2}{q_0^2}\;\hat N \; R^2 N_F\left(r_0\right) {\cal P}_T(P) \Sigma_{\rm HTL}(d)(R), \nonumber\\
\text{Re}\Sigma(d_T,{\cal P}_+,N_B) & = & -\frac{1}{q_0^2} \;\hat N \;\left(p_0-r\right)^2 \left(r+p_0+2 q_0\right)^2 d_T(P)
   N_B\left(p_0\right) {\cal P}_{+}(R),\nonumber\\
\text{Re}\Sigma(d_T,{\cal P}_-,N_B) & = &-\frac{1}{q_0^2} \;\hat N \;\left(r+p_0\right)^2 \left(-r+p_0+2 q_0\right)^2 d_T(P)
   N_B\left(p_0\right) {\cal P}_{-}(R),\nonumber\\
\text{Re}\Sigma(d_+,{\cal P}_T,N_F) & = &-\frac{1}{q_0^2} \;\hat N \;\left(p_0-r\right)^2 \left(r+p_0+2 q_0\right)^2 d_{+}(R)
   N_F\left(r_0\right) {\cal P}_T(P),\nonumber\\
\text{Re}\Sigma(d_-,{\cal P}_T,N_F) & = &-\frac{1}{q_0^2}\;\hat N \; \left(r+p_0\right)^2 \left(-r+p_0+2 q_0\right)^2
   d_{-}(R) N_F\left(r_0\right) {\cal P}_T(P). \nonumber
\eea
The sixth equation in (\ref{real-part}) contains the part of the lowest order result that comes from the Bose-Einstein distribution function and the last two equations contain the part of the lowest order result that comes from the Fermi-Dirac distribution function. The results for these two terms are
\bea
\text{Re}(d_T,N_B) & = &\frac{e^2T^2}{12}+\cdots, \nonumber\\
\text{Re}(d_+,{\cal P}_T,N_F)+\text{Re}(d_-,{\cal P}_T,N_F) & = &\frac{e^2T^2}{24}+\cdots
\eea
where the dots indicate the NLO contribution.

\section{Numerical analysis}
\label{sec:num}
\begin{figure}[H]
\begin{center}
\includegraphics[width=10cm]{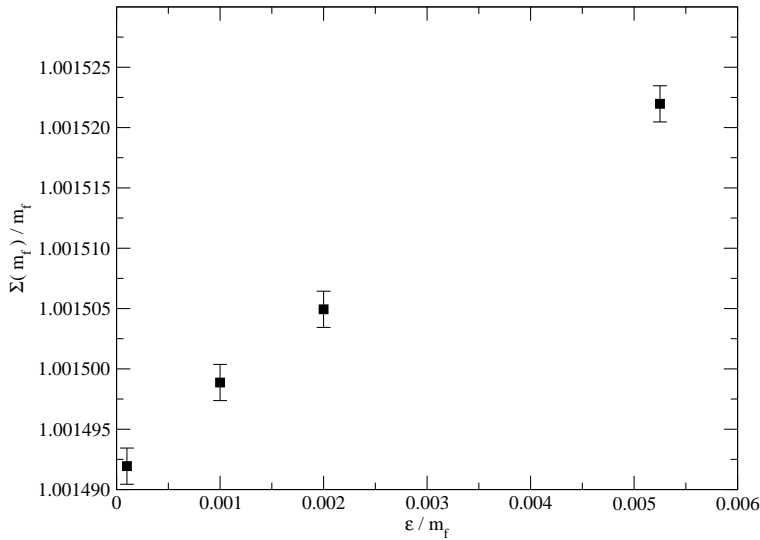}
\end{center}
\caption{Extrapolation of the numeric estimate of the sum of integrals given in Eq.~\ref{real-part} to $\epsilon \rightarrow 0$ as defined in the text (for QCD at $g=0.005$).} 
\label{extrapolation}
\end{figure}

In this section we briefly describe the numeric methods used to get the results given in Eq.~(\ref{result}).
The dimensionless integrals to be evaluated numerically, given in Eq.~(\ref{real-part}), are of the form
\begin{equation}
I(g) = \int_{0}^{\infty}\!\!dp \int_{-\infty}^\infty\!\! dp_0 \,f(g,p_0,p)\,,
\end{equation}
where $g$ is the coupling constant.  The integrand $f(g,p_0,p)$ diverges along some curve $p_0=h(p)$ like $[p_0-h(p)]^{-1}$ and therefore  the integrals must be defined using a principle value prescription,
\begin{equation}
I(g) = \lim_{\epsilon\rightarrow 0}\left[ \int_0^\infty \!\! dp \left(\int_{-\infty}^{h(p)-\epsilon}\!\!\!\!dp_0 \,f(g,p_0,p) + \int_{h(p)+\epsilon}^\infty \!\!\!\! dp_0 \,f(g,p_0,p) \right) \right]. 
\end{equation}
The curve $h(p)$ can be computed numerically for each integrand to a high and controllable accuracy. We write  $(\tilde{h}(p)-h(p))/h(p) \sim \delta$, where $\tilde{h}(p)$ is the numeric estimate of $h(p)$ and $\delta$ is a measure of the error. A numeric approximation of the required integrals is obtained as
\begin{equation}
\tilde{I}(g,\epsilon) = \int_0^\infty \!\! dp \left(\int_{-\infty}^{\tilde{h}(p)-\epsilon}\!\!\!\!dp_0 \,f(g,p_0,p) + \int_{\tilde{h}(p)+\epsilon}^\infty \!\!\!\! dp_0 \,f(g,p_0,p) \right) + \mathcal{O}(\epsilon).
\end{equation}   
The integral must be evaluated for a number of different values of $\epsilon$ and then extrapolated to $\epsilon \rightarrow 0$. However, for numeric stability, one must require $\epsilon \gg \delta\, \tilde{h}(p)$. As a consequence, we have to estimate $h(p)$ to very high accuracy in order to get a reliable extrapolation of the limit $\epsilon \rightarrow 0$. This extrapolation is illustrated in Fig.~\ref{extrapolation} for one value of the coupling constant.

\vspace*{.5cm}

%%%%%%%%%%%%%%%%%%%
\par
%%%%%%%%%%%%%%%%%%%%

\begin{figure}[H]
\begin{center}
\includegraphics[width=10cm]{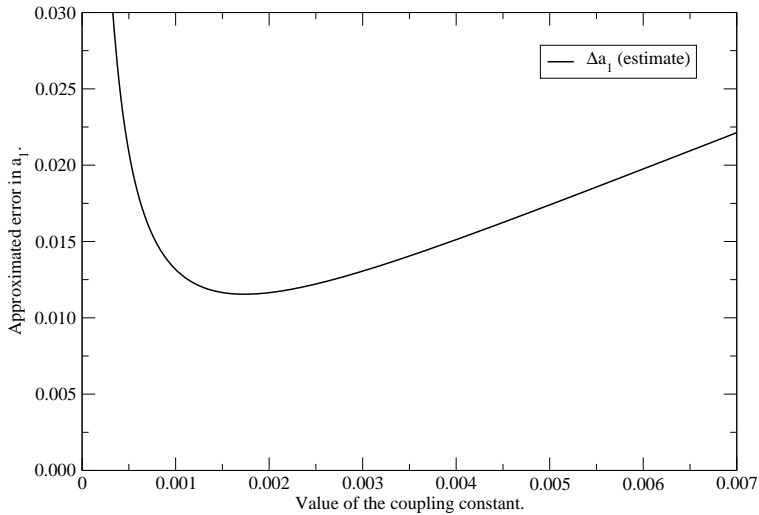}
\end{center}
\caption{The error in $a_1$ as a function of the coupling constant.}
\label{coupling}
\end{figure}

In order to extract the NLO correction to the thermal fermion mass we must further extrapolate the result to small values of the coupling constant. Using the LO result to set the dimensions, we use the ansatz
\begin{equation}
\label{ansatz}
\frac{\Sigma(m_f)}{m_f} = 1 + a_1 \frac{g}{2\pi} + a_2 \left(\frac{g}{2\pi}\right)^2\left(\ln 1/g + a_2'\right) + \mathcal{O}(g^3),
\end{equation}
where we assume that all coefficients $a_i$ are of order one. Our goal is to determine the coefficient $a_1$. The accuracy of the result depends on two things:
the accuracy of the numeric estimate of the integrals, and the size of the error that is made by neglecting higher order terms from the ansatz. If we drop terms of order $g^2$, the error in $a_1$ can be estimated as
\begin{equation}
\label{error}
\Delta a_1 \approx \left|\frac{2\pi}{g}\frac{\Delta\Sigma}{m_f}\right| + \left|a_2 \frac{g}{2\pi}(\ln(1/g)+a_2')\right|,
\end{equation}
where $\Delta\Sigma$ is the error in $\Sigma(m_f)$ from the numeric estimation of the integrals. 
Minimizing the error in $a_1$ determines the optimal range of values of the coupling constant at which the integrals should be computed. We are able to obtain an accuracy of the order of $\Delta\Sigma/m_f \approx 10^{-6}$ which means that using $0.001 \lesssim g \lesssim 0.006$ we obtain an error in the result for $a_1$ of order $\Delta a_1 \approx 0.02$ (see Fig.~\ref{coupling}).

\vspace*{.8cm}

%%%%%%%%%%%%%%%%%%%
\par
%%%%%%%%%%%%%%%%%%%%

\begin{figure}[H]
\begin{center}
\includegraphics[width=10cm]{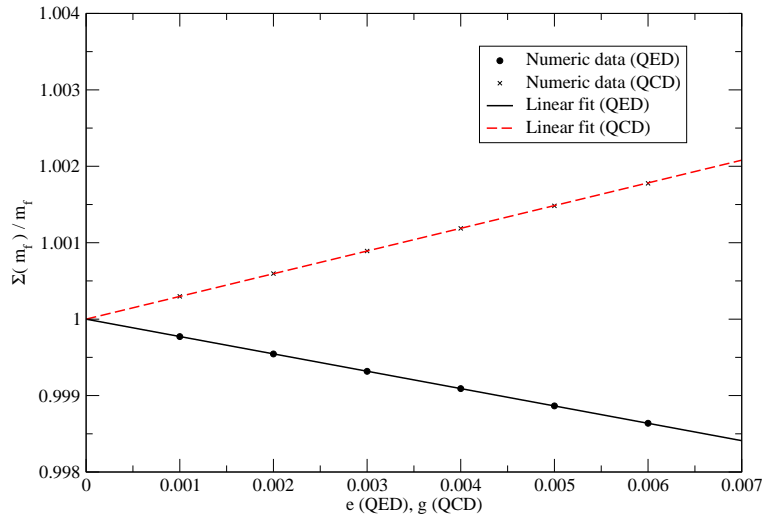}
\end{center}
\caption{The numeric results for $\Sigma(m_f)/m_f$ along with the best fit curve.}
\label{num-result1} 
\end{figure}

\begin{figure}[H]
\begin{center}
\includegraphics[width=10cm]{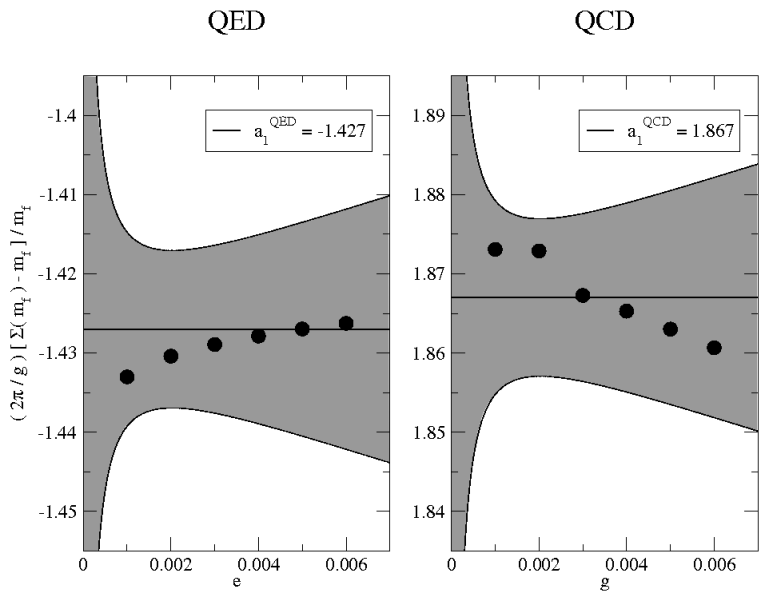}
\end{center}
\caption{The numeric results for $2\pi/g (\Sigma(m_f)-m_f)/m_f$. The dots are the result of the numeric computations, the solid lines correspond to the numeric estimate for $a_1$ and the gray regions correspond to the estimated errors as defined in Eq.~(\ref{error}).} 
\label{num-result2}
\end{figure}

In Figs.~\ref{num-result1} and \ref{num-result2} we have plotted the numeric results for $\Sigma(m_f)/m_f$ and $(\Sigma(m_f)-m_f)/m_f\cdot 2\pi/g = a_1 + \mathcal{O}(g)$ along with the best fit (least-square) curve. The best fit results for $a_1$ are
\begin{eqnarray}
a_1^\mathrm{QED} & = & -1.427 \pm 0.02, \\
a_1^\mathrm{QCD} & = & 1.867 \pm 0.02 \hspace{1cm} (N_f = 2,\, N_c = 3). \nonumber
\end{eqnarray}  

\vspace*{.8cm}

%%%%%%%%%%%%%%%%%%%
\par
%%%%%%%%%%%%%%%%%%%%

%%%%%%%%%%%%%%%%%%%
\par
%%%%%%%%%%%%%%%%%%%%

\section{Discussion and Conclusions}
\label{conc}

The calculation of dispersion relations for soft quantities at next-to-leading order in the hard thermal loop effective theory is extremely difficult. Real quantities are particularly hard to obtain. 
Our results for the next-to-leading fermion mass in QED and QCD are given in Eqn (\ref{result}). 
%%%\bea
%%%&&M_{\rm QED} =\frac{e T}{\sqrt{8}}\left(1 - (1.417\pm 0.01)\frac{e}{4\pi}\right) +\mathcal{O}(e^3T) \\[2mm]
%%%&&M_{\rm QCD} = \sqrt{C_F}\;\frac{g T}{\sqrt{8}}\left(1 + (1.883\pm 0.01)\frac{g}{4\pi}\right) + \mathcal{O}(g^3T) \nonumber
%%%\eea
%%%The term of order $e(eT)$ (or $g(gT)$) is the full NLO contribution. The term of order $e^3\ln(1/e)T$ (or $g^3\ln(1/g)T$) is gauge independent, which suggests that it might be the complete contribution at this order. 

It was pointed out in \cite{mitra} that the subleading correction to the 1-loop HTL contribution (calculated by expanding to next  order in the ratio of the soft external momentum to the hard loop momentum) is gauge dependent and of order $\sim e^3 \ln(1/e)\,T$. There is a contribution of the same order  from 2-loop diagrams where one loop momentum is of order $T$ and the other loop momentum contributes a log term coming from an integral of the form $\int dp \frac{1}{p} \to \ln(p_{\rm hard}/p_{\rm soft}) \sim \ln(1/e)$. In \cite{emil} it was shown that the sum of these two contributions is gauge independent (in the class of covariant gauges). 
Both of these contributions contribute to the coefficient $a_2$ in Eqn. (\ref{ansatz}) and are formally included in the integrals given in this paper, which correspond to the diagrams in Fig. \ref{KKMdiag2}. 
The gauge independence of the result in \cite{emil} suggests that the full contribution at order $g^3 \ln\,1/g$ might be contained in these diagrams.

\end{document}